\documentclass[12pt]{article}
\usepackage{amsmath,epsfig}

\usepackage{amsmath,amsthm,amsfonts,pmat}
\usepackage{color}
\usepackage{graphicx}

 \def\BibTeX{{\rm B\kern-.05em{\sc i\kern-.025em b}\kern-.08em
     T\kern-.1667em\lower.7ex\hbox{E}\kern-.125emX}}
\newtheorem{theorem}{Theorem}

\newtheorem{corollary}{Corollary}

\def\ve{{\bf e}}

  \def\vs{{\bf s}} \def\vt{{\bf t}}
 \def\vv{{\bf v}}

\def\vA{{\bf A}}  \def\vC{{\bf C}} \def\vD{{\bf D}}
   
\def\vI{{\bf I}}   
   
 \def\vR{{\bf R}}

\def\v0{{\bf 0}}

\def\lam{{\mbox{\boldmath $\Gamma$}}}

\def\bxi{{\mbox{\boldmath $\xi$}}}

\def\bxi{{\mbox{\boldmath $\xi$}}}

\def\ie{{\it i.e.}}

\begin{document}

% "Title of the paper"
\title{The Rank of the Covariance Matrix of an Evanescent Field}

\author{Mark  Kliger and Joseph M. Francos
%\thanks{M.  Kliger is with Medasense Biometrics Ltd., PO Box 633, Ofakim 87516, Israel. Tel: +972 8 9921182
%FAX: +972 8 9926581, email: mark@medasense.com. \newline
%J. M.~Francos is with the Department of Electrical and Computer
%Engineering, Ben-Gurion University, Beer-Sheva 84105, Israel. Tel:
%+972 8 6461842, FAX: +972 8 6472949, email: francos@ee.bgu.ac.il}
}

\nopagebreak
 \maketitle
 \nopagebreak

\begin{abstract} 
Evanescent random fields  arise as a component of the 2-D Wold decomposition of homogenous random fields. Besides their theoretical
importance, evanescent random fields have a number of practical applications, such as in modeling the observed signal in the space time adaptive processing (STAP) of airborne radar data.
In this paper we derive an expression for the rank of the low-rank covariance matrix of a finite dimension sample
from an evanescent random field.  It is shown that the rank of this covariance matrix is completely determined by the evanescent field
spectral support parameters, alone. Thus, the problem of estimating the rank lends itself to a solution that avoids the need to estimate the
rank from the sample covariance matrix. We show that this result can be immediately applied to considerably simplify the estimation of the rank of the  interference covariance matrix in  the STAP problem.
\end{abstract}

%{ \bf Keywords:} [class=AMS]
%\kwd[Primary ]{60G60}
%\kwd[; secondary ]{62M20}
%\kwd{62M40}
%\kwd{60G35}
%
%
%\begin{keyword}
%\kwd{Homogeneous random fields}
%\kwd{evanescent random fields}
%\kwd{covariance matrix}
%\kwd{linear Diophantine equation}
%\end{keyword}
\vspace{.5in}


\pagestyle{plain}
\newpage

%%%%%%%%%%%%%%%%%%%%%%%%%%%%%%%%%%%%%%%%%%%%%%%%%%%%%%
%
%       Introduction
%
%%%%%%%%%%%%%%%%%%%%%%%%%%%%%%%%%%%%%%%%%%%%%%%%%%%%%%

\section{Introduction}\label{Intro}
\subsection{The Evanescent Random field}
The problem of linear prediction of stationary processes is a classic problem in time-series analysis. One of the most fundamental results in this field is  the Wold decomposition \cite{wold1}, that states that a regular one dimensional wide-sense stationary processes indexed by ${\mathbb Z}$ may be decomposed into two stationary and orthogonal components: the purely-indeterministic process (that produces the innovations) and the deterministic process. This decomposition can be equivalently reformulated using spectral notations: the spectral measure of the purely-indeterministic process is absolutely continuous with respect to the Lebesgue measure, and the spectral measure of the deterministic process is singular. In other words, the spectral measures of the orthogonal components of Wold decomposition yield the Lebesgue decomposition of the spectral measure of the process.

\begin{figure}
\begin{center}
  \includegraphics[width=3in]{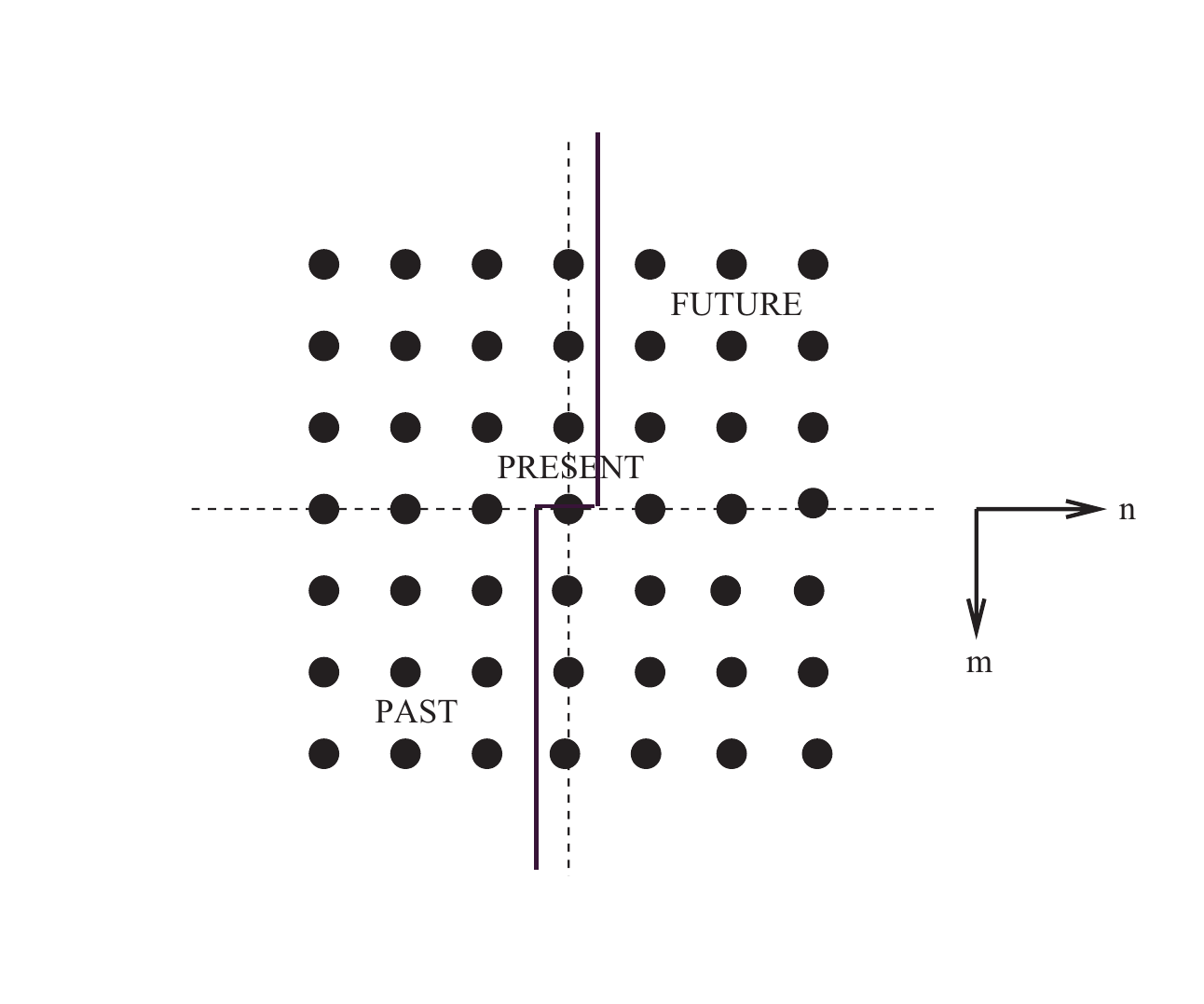}\\ %nshp_1.pdf
  \end{center}
\caption{Non-symmetric half plane }\label{intro_f}
\end{figure}

Homogenous random fields, (also called doubly stationary series), are the two-dimensional (indexed by ${\mathbb Z}^2$) generalization of one-dimensional wide-sense stationary process. Unfortunately, unlike the one-dimensional case, in multiple dimensions there is no natural order definition and  terms such as ``past" and ``future" are meaningless unless defined with respect to a specific order. Linear prediction of homogenous random fields was first rigorously formulated by Helson and Lowdenslager in \cite{Helson1}. The problem of defining ``past" and ``future" on the two-dimensional lattice (i.e., ${\mathbb Z}^2$) was resolved in \cite{Helson1} in terms of ``half plane" total-ordering. The trivial example of a half-plane total order on $\mathbb Z ^2$ is a usual lexicographic order:  $(k,l) \preceq (n,m)$ \emph{iff} $k < n$ or ($k = n$ and $l < m$). Lexicographic order can be considered as a linear order induced by Non-Symmetric (delimited by a broken straight line) Half Plane (NSHP), (see Figure \ref{intro_f}).

Further analysis of the prediction problem led to a generalization of the Wold decomposition \cite{Helson2}.  When we consider random processes indexed by a group we obtain a Wold decomposition with respect to any given total order on the group. When the group is not $\mathbb Z$ (like $\mathbb R$ or $\mathbb Z ^2$) the deterministic process can have as a direct summand a deterministic process of a special type, the \emph{evanescent process}.  In order to provide some intuition on the characteristics of the evanescent process we next state some basic definitions and present an example of an evanescent field defined with respect to a vertical total order, which is simply a lexicographic order on $\mathbb Z ^2$:

A homogeneous random field $\{y(n,m)\}$
is called {\it regular} with respect to the {\it lexicographic order} if
for every $(n,m)$,  $E[y(n,m)-\hat y(n,m)]^2=\sigma^2 >0$ where $\hat
y(n,m)$ is the projection of $y(n,m) $ on the $ c.l.m.\, \Big[
\{y(k,l):\, k<n, l\in\mathbb Z\}\cup \{ y(n,l): \, l< m\}\Big]$, where $c.l.m.$
denote a closed linear manifold. Thus, a regular homogeneous random field  has a non-zero innovation at every lattice point. A homogeneous random field $\{z(n,m)\}$ is called {\it deterministic} with respect to the lexicographic order if it can be perfectly linearly predicted from its past in mean-square sense, \ie, for
every $(n,m)$ we have $z(n,m) \in  c.l.m.\, \Big[ {\{z(k,l):\, k<n,
l\in\mathbb{Z}\}}\cup\{ z(n,l): \, l< m\}\Big] $.

Although a deterministic field can be perfectly predicted from its past with respect to lexicographic order, it may still  posses a non-zero innovation when  prediction is based on samples  in previous columns only. We then say  that the field $\{z(n,m)\}$
has vertical {\it column-to-column innovations} if
$I(n,m):=z(n,m)-\hat z(n,m)$ (the {\it innovation}) is not 0, where
$\hat z(n,m)$ is the orthogonal projection of $z(n,m)$ on the closed
subspace generated by $\{z(k,l):\, k<n, l\in\mathbb{Z}\}$. In other words, if a deterministic field has  non-zero column-to-column innovations it cannot be perfectly linearly predicted from previous columns.

When $z(n,m)$ is the deterministic component of the decomposition of a regular random field with respect to a NSHP total-ordering, the vertical evanescent component
$z_e(n,m)$ is the orthogonal projection of $z(n,m)$ on the closed
subspace generated by the (orthogonal) column-to-column innovations
$\{I(k,l):\, k\le n, l\in \mathbb{Z}\}$. Thus, an evanescent field spans a Hilbert space identical to the one spanned by column-to-column innovations.  In other words, the evanescent field is a component of the deterministic field which represents column-to-column innovations. Horizontal column-to-column (row-to-row) innovations and evanescent components are similarly defined.

\begin{figure}
\begin{center}
  \includegraphics[width=3in]{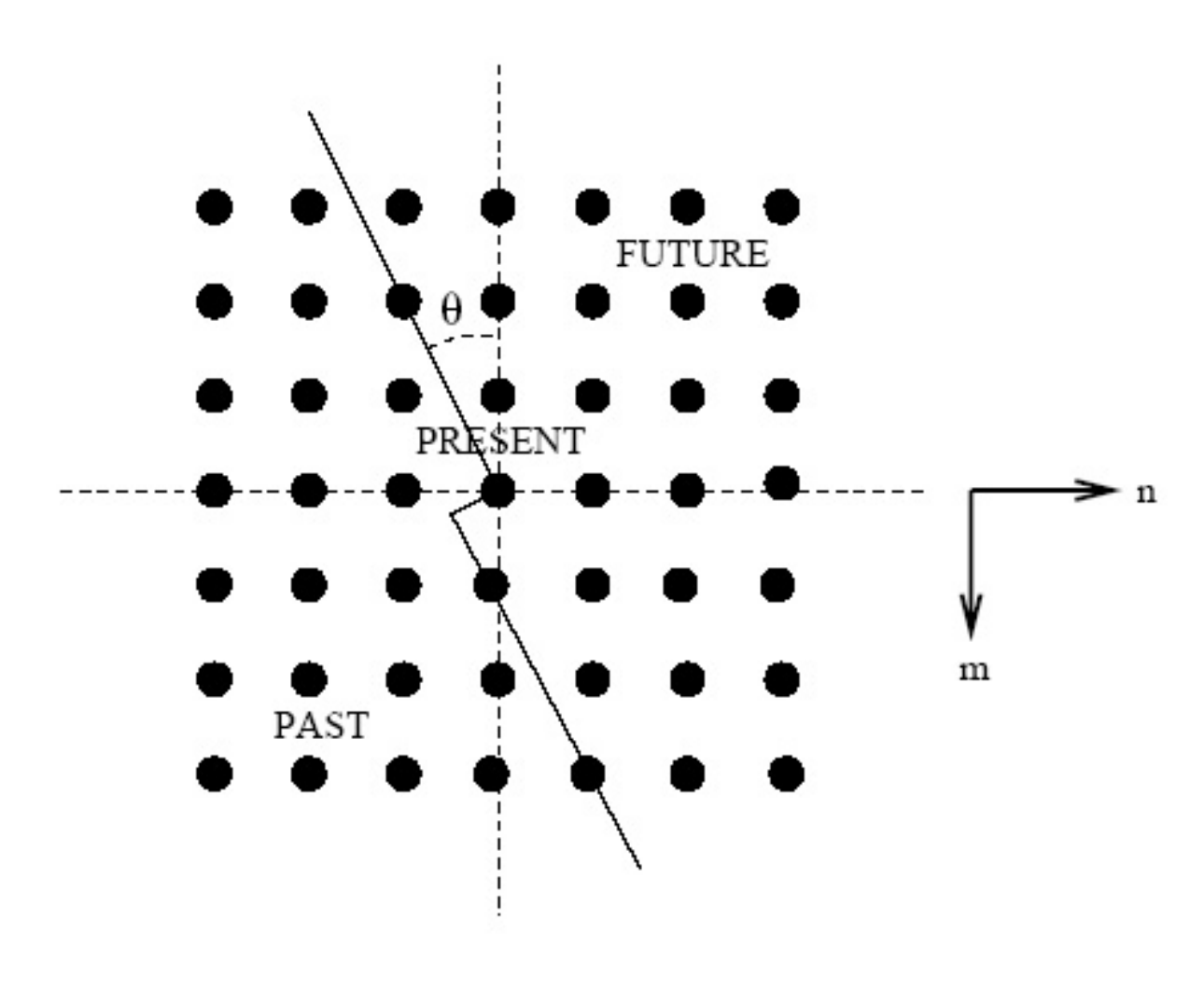}\\
  \end{center}
\caption{RNSHP support.}\label{intro_f_1}
\end{figure}

Evanescent
processes were first introduced in \cite{Helson2} (on ${\mathbb R}$).
In Korezlioglu and Loubaton \cite{korez}, ``horizontal" and ``vertical"
total-orders and the corresponding horizontally and vertically evanescent
components of a homogeneous random field on ${\mathbb Z}^2$ are defined.
In Kallianpur \cite{kall},
as well as in Chiang \cite{chiang}, similar techniques are employed to obtain
four-fold orthogonal decompositions of regular (non-deterministic) homogeneous
random fields.
In Francos {\it et. al.} \cite{Francos1} this  decomposition  of
random fields on ${\mathbb Z}^2$ was further extended. This is
done by considering {\it all} the Rational Non-Symmetrical Half Plane (RNSHP) linear
orders, each inducing a different partitioning of  the
two-dimensional lattice into two sets by a  broken straight line of
rational slope. Intuitively, the usual lexicographic order is not the only possible order definition of the 2-D lattice. Each RNSHP linear order is induced by a ``rotation" of the usual lexicographic order, such that the resulting non-symmetrical half-plane is delimited by a broken straight line with rational slope, and which leads to a different linear order definition. Consequently, terms such as ``past" and ``future" are redefined with respect to a specific RNSHP linear order (see, for example, Figure \ref{intro_f_1}).

 More specifically, each
Rational Non Symmetrical Half Plane is defined in terms of  two
co-prime integers $(a,b)$, such that  the past $P_{a,b}$ is defined by
\begin{equation}
\label{porder} P_{a,b}= \{(n,m)\in\mathbb Z^2: na+mb<0\ \mbox{, or}\
na+mb=0\ \mbox{and}\ m\le0\}.
\end{equation}
Then $P=P_{a,b}$ satisfies
$$
(i)\  P\cap (-P)=\{0\},\quad (ii)\  P\cup (-P)=\mathbb Z^2,\quad
(iii)\  P+P\subset P\ (usual\  addition).
$$
By (i)-(iii), $P$ induces on $\mathbb Z^2$ a linear order, which is
defined by $(k,l)\preceq (n,m)$ if and only if $(k-n,l-m)\in P$.

Clearly, there are countably many such linear orders. Each such order  induces a different definition of the term ``column", and correspondingly   different definitions of column-to-column innovations and evanescent field.

The Wold decomposition of a regular random field into purely-indeterministic
and deterministic components is invariant to the choice of a RNSHP order. The decomposition in \cite{Francos1} further asserts that we can represent the
deterministic component of the field as a mutually orthogonal sum of a
``half-plane deterministic" field  and a countable number of evanescent fields.
The half-plane deterministic field has no innovations, nor
column-to-column innovations, with respect to any
RNSHP linear order. On the other hand, each of the evanescent fields can be
revealed only by using the corresponding RNSHP linear order, \ie, with respect to specific definitions of ``columns" and column-to-column innovations.
This decomposition yields a corresponding spectral decomposition,
\ie, we can decompose  the spectral measure of the deterministic
part into a countable sum of mutually singular spectral measures, such that the spectral measure
of each evanescent component is concentrated on a line with a rational slope.

Based on these results, a  parametric model of the homogeneous random field was derived
in \cite{Francos1}. The purely-indeterministic component of the field is modeled by a white innovations driven 2-D moving average  process with respect to some RNSHP linear order. This component contributes the absolutely continuous part of the spectral measure of the regular field. One of the components of the half-plane deterministic component that is often found in practical applications is  the 2-D harmonic random field which is the sum of a countable number of exponential components, each having a constant spatial frequency and random amplitude. This component contributes the 2-D delta functions in the spectral domain. The number of evanescent components of the regular field is countable. The model of the evanescent field with respect to specific order is presented bellow:

Let $(a,b)$ be a pair of co-prime integers ($a\geq 0$) which defines a specific RNSHP linear order according to (\ref{porder}). Then, the model of the evanescent field which corresponds to this order is
\begin{eqnarray}
e_{(a,b)}(n, m)=
\sum_{i=1}^{I_{(a,b)}} s_i^{(a,b)}(n a+m b)
\exp \biggl ( {j {\omega^{(a,b)}_{i}}
(n c +m d)} \biggr )   \ ,
\label{e30}
\end{eqnarray}
where $c$ and $d$ are co-prime integers satisfying $ad-bc=1$. For the case where $(a,b)=(0,1)$ we have $(c,d)=(1,0)$, and for $(a,b)=(1,0)$ we have $(c,d)=(0,1)$.  We further note that in this notation $na+mb$ is the ``column" index and $nc+md$ defines a ``row". The modulating process $ \{ s_i
^{(a,b)}(n a+m b) \} $ is a 1-D purely-indeterministic, complex valued processes, and $\omega^{(a,b)}_{i}$ is a modulation frequency.  Thus, $e_{(a,b)}$ has no innovations, with respect to ``rows", and has non-zero column-to-column innovation (expressed by the modulating process $s_i^{(a,b)}$) with respect to  its ``columns".  $I_{(a,b)}$ denotes the number of different evanescent components
that correspond to the same RNSHP defined by $(a,b)$.
The different components are such that their 1-D modulating processes $ \{ s_i
^{(a,b)} \} $ and $ \{ s_j ^{(a,b)}\} $, are {\it mutually orthogonal}
and their modulation frequencies are different
$\omega^{(a,b)}_{i} \neq \omega^{(a,b)}_{j}$ for all $1\leq i \ne j \leq I_{(a,b)}$.

The ``spectral density function" of each
evanescent field has the form of a sum of 1-D delta functions which
are supported on lines of rational slope in the 2-D spectral domain. The amplitude of each of these
delta functions is determined by the spectral density of the 1-D modulating process. Since the
spectral density of the modulating process can rapidly decay
to zero, so will the ``spectral density" of the evanescent field, and hence the name ``evanescent".

\subsection{ Practical Applications}
 Besides its  fundamental theoretical
importance, the Wold decomposition of a regular random field has
various applications in image processing and wave propagation
problems. For example, the parametric model that results from these
orthogonal decompositions, naturally arises as the physical model in
problems of texture modeling, estimation and synthesis
\cite{francos2}.

Another application is space-time adaptive processing of airborne radar data \cite{STAP}. Space-time adaptive processing (STAP) is an increasingly popular radar signal processing technique for detecting slow-moving targets. The space dimension arises from the use of array of multiple antenna elements and the time dimension arises from the use of coherent train of radar pulses. The power of STAP comes from the joint processing along the space and time dimensions. Comprehensive analysis of STAP appears in \cite{klemm,ward}.

In \cite{STAP} it is shown that the same parametric model that results from the 2-D Wold-like orthogonal decomposition naturally arises as the physical model in the problem of space-time processing of airborne radar data. This correspondence  is exploited to derive  computationally efficient %fully adaptive and partially adaptive
detection algorithms.
More specifically, the target signal is modeled as a random amplitude complex exponential where the exponential is defined by a space-time steering vector that has the target's angle and Doppler.
Thus, in the space-time domain the target
contribution is the half-plan deterministic component of the
observed field. The sum of the white noise field
due to the internally generated receiver amplifier noise, and the
sky noise contribution, is the purely-indeterministic component of
the space-time field decomposition.

The  presence of a jammer (a foe interference source, transmitting high power noise
aimed at ``blinding" the radar system) results in a barrage of noise localized in angle
and uniformly distributed over all Doppler frequencies (since the transmitted noise is white). Hence, in the space-time
domain each jammer is modeled as an evanescent component with $(a,b)=(0,1)$ such that its 1-D modulating
process $s_i^{(0,1)}(m)$ is the random process of the jammer amplitudes. The jammer samples from different pulses are uncorrelated.
In the angle-Doppler domain each jammer
contributes a 1-D delta function, parallel to the Doppler axis and
located at a specific angle $\omega_i^{(0,1)}$ using the notation of (\ref{e30}).

The ground clutter results in an additional
evanescent component of the observed 2-D
space-time field. The aircraft platform motion produces a very special structure of the clutter due to the dependence of the Doppler frequency on angle. The clutter's echo from a single ground patch
has a Doppler frequency that linearly depends on its aspect with respect to the platform. As the platform moves,  identical clutter observations are repeated by different antenna elements on different pulses, which defines a specific linear locus in the angle-Doppler domain, commonly referred as the ``clutter ridge".  Thus, the clutter ridge, which represents clutter from all angles, is supported on a diagonal line (that generally wraps around) in the angle-Doppler domain. Due to the physical properties of the problem the different components of the field are assumed to be mutually orthogonal. In the specific application of airborne radar, the evanescent components (the clutter, and jamming signals) are considered unknown interferences.

Although the data collected by STAP radars for different ranges can be viewed as a sequence of finite-sample realizations from a homogenous field, its is technically more convenient to represent each of the observations in a vector form and to statistically analyze them as  multivariate vectors. Thus, if one  uses a STAP system with $N$ antenna elements and $M$ pulses, the observed $N \times M$  STAP signal is treated as $NM \times 1$ multivariate random variable. These vectors are commonly called ``snapshots".

The STAP processor goal is to solve a detection problem, \ie, to establish whether a hypothetical target is present or not. It adaptively weights the available data in order to achieve high gain at the target's angle
and Doppler and maximal mitigation along both the jamming and clutter lines. The adaptive weight vector is computed from the inverse of the interference-plus-noise covariance matrix,\cite{klemm,ward}. It is shown in
\cite{haim} that the dominant eigenvectors of the space-time covariance matrix contain all the information
required to mitigate the interference. Thus, the weight vector is constrained to be in the subspace orthogonal to the
dominant eigenvectors. Because the interference-plus-noise covariance matrix is unknown a priori, it
is typically estimated using  sample covariances obtained from averaging over a few range gates.
This is the known as the {\it fully adaptive} STAP approach. The major drawback of this approach is its high computational complexity. The final detection of a target is performed by applying either Constant-False-Alarm-Rate (CFAR) detector, or Adaptive-Matched-Filter (AMF) detector, or Generalized-Likelihood-Ratio (GLR) detector. Usually the detector is embedded into the weight computations.

Fortunately, both the clutter and the jammers have low-rank covariance matrices. The clutter covariance matrix has a low rank due to the movement of the platform, as discussed above.  The jammer covariance matrix has low rank since the jamming signal is spatially correlated between all antennas at each pulse. The {low-rank} structure of the interference covariance matrix may be exploited to achieve significant reduction in the adaptive problem dimensionality with little or no sacrifice in performance relative to the fully adaptive case. These methods are referred to as {\it partially adaptive} STAP.
%In \cite{reed}, a  reduced rank CFAR
%detection test is developed, assuming
%the dominant eigenvectors of the interference are known. In \cite{gold} a multistage {\it partially adaptive}
%CFAR detection algorithm is introduced. In  \cite{sarkar2000}, an approach that bypasses
%the need to estimate the covariance matrix is presented: The data collected in a single range gate
%is employed to obtain a least-squares estimate of the signal power at each hypothesized direction of arrival, through evaluation of a weight vector constrained to null the unknown interference and noise. In \cite{wang}, a simple ad-hoc model of the clutter signal and covariance matrix is proposed. The model represents the spectral density of the clutter as a sum of Gaussian-shaped humps along the support of the clutter ridge. In \cite{blum}, this model is employed to estimate the clutter covariance matrix from the data observed in a single range gate. Finally, in \cite{STAP} fully and partially adaptive methods has been derived which exploit a parametric model of airborne data based on Wold decomposition.

Partially adaptive STAP methods  require  knowledge of the rank of the interference covariance matrix. However, it is  a priori unknown, and unfortunately cannot be easily estimated from the sample covariance matrix  due to the existence of a noise component which has a full rank covariance matrix. Hence, the problem of estimating the rank of the interference covariance matrix is critical in the implementation of  many STAP algorithms.

In this work we consider the problem of determining the rank of the
covariance matrix of a vectorized finite dimension sample from an evanescent
random field. By using the evanescent field parametric model to model the  interferences in the STAP problem, we considerably simplify the solution to  the problem of estimating the rank of the low-rank interference covariance matrix. In fact, it turns out that in this parametric framework the well-known Brennan rule \cite{ward} for the  rank of the clutter covariance matrix, as well as the rank computation of the  jammer, become special cases of the
general result proved here. %Its proof, however, employs an entirely different
%approach than the model based geometrical-combinatorial method
%described this paper.
Hence, the provided derivation opens the
way for new, computationally attractive, methods in parametric and
non-parametric estimation of two-dimensional random fields, with
immediate applications in partially adaptive space-time
adaptive processing of airborne radar data.

The rest of the paper is organized as follows: In Section \ref{notations} we formulate the problem we aim to solve. A formula for the rank of the
covariance matrix of a complex valued evanescent random field is
derived in Section \ref{comp_rank}. In Section \ref{real_rank} we
extend the obtained result to the case of a real valued evanescent
random field. Finally, in Section \ref{Concl} we provide our
conclusions.

The following notation is used throughout. Boldface upper case letters denote
matrices, boldface lower case letters denote column vectors, and
standard lower case letters denote scalars. The superscripts
$(\cdot)^T$ and $(\cdot)^{H}$ denote the transpose and Hermitian transpose operators, respectively.
By $\vI$ we denote the identity matrix and by $\v0$ a matrix of zeros. The symbol $\odot$ denotes an element by element product of the vectors
(Hadamard product). Given a scalar function $f(\cdot)$ and a column vector $\vv$, we denote by
$f(\vv)$ a column
vector consisting of the values of function $f(\cdot)$ evaluated for all the
elements of $\vv$. Finally $diag(\vv)$
denotes a square diagonal matrix with the elements of $\vv$ on its main
diagonal.

%%%%%%%%%%%%%%%%%%%%%%%%%%%%%%%%%%%%%%%%%%%%%%%%%%%%%%
%
%       Notation and Definitions
%
%%%%%%%%%%%%%%%%%%%%%%%%%%%%%%%%%%%%%%%%%%%%%%%%%%%%%%

\section{Finite Sample of an Evanescent Random Field: Definitions and Problem Formulation}
\label{notations}

Let  $O$ denote the set of all possible pairs of different co-prime integers $(a,b)$, $a\ge 0$, where each pair defines a RNSHP order on the 2-D lattice. Although in the case of an infinite 2-D lattice the number of different RNSHP definitions is infinitely countable, in the finite sample case only a finite number of different linear orders can be defined. Moreover, in practical applications the number of different evanescent components for each order definition is finite as well. Therefore, we assume throughout this paper that $|O|$ and $I_{(a,b)}$ are finite integers.

Let $Q=\sum_{(a,b)\in O}I_{(a,b)}$. Each of the $Q$ evanescent components is uniquely defined by the triple $(a,b,\omega^{(a,b)}_{i})$ where $(a,b)\in O$ and $\omega^{(a,b)}_{i}$ is the frequency parameter. Let us denote the set of all possible triples by $O_Q=\{(a_1,b_1, \omega_{1}),(a_2,b_2,\omega_{2}),\ldots, (a_Q,b_Q,\omega_{Q})\}$. All triples are unique: they either have different support parameters $(a_i,b_i)\neq(a_j,b_j)$, or in case $(a_i,b_i)=(a_j,b_j)$ they have different  frequencies such that  $\omega_{i}\neq \omega_{j}$.

 Finally, adapting (\ref{e30}) to the finite sample case we have that
\begin{equation}
e(n,m)= \sum_{q=1}^{Q}e_q(n,m), \label{eq1a}
\end{equation}
where
\begin{equation}
e_q(n,m)=s_q(n a_q+m b_q) \exp \biggl ( {j {\omega_q} (n c_q +m d_q)} \biggr ) \label{eq1b}
\end{equation}
such that $(a_q,b_q,\omega_q) \in O_Q$ and $\{s_q\},c_q,d_q$ are defined as above.

We note that since the spectral
measure of  $\{e_q(n, m)\}$ is concentrated on a line (that
may wrap around) whose slope is determined by $a_q$ and $b_q$, we
interchangeably  refer to $a_q$ and $b_q$ as  either  the {\it spectral support parameters} of
$\{e_q\}$ or as the {\it RNSHP slope parameters}.

Let $\lbrace e(n,m): (n,m) \in D \rbrace$
where $D = \{(n,m)\in
\mathbb{Z}^2: 0 \le n \le N - 1,\,0 \le m \le M - 1\} $ be the observed finite sample of the  random field (\ref{eq1a}). Let $\ve$ denote an $NM \times 1$ vector form representation of this finite sample:
\begin{eqnarray}
{\bf e}&=&\left [
e(0,0),\dots,e(0,M-1),e(1,0),\dots,e(1,M-1),\dots, \right. \nonumber \\
&&  \left.  \dots, e(N-1, 0) , \dots, e(N-1,M-1) \right ]^T.
\label{c16-1}
\end{eqnarray}
This is a multivariate representation of a finite sample of an evanescent random field.

Let $\lam$ denote the $NM \times NM$ covariance matrix of the evanescent vector $\ve$,
\begin{equation}
\lam= E \bigg [\ve \left (\ve\right )^H \bigg ].
\end{equation}

Due to the special structure of the evanescent field, many of the elements of $\ve$ are linearly dependent, and therefore $\lam$ is low-rank. This property is easily demonstrated by considering a single evanescent component that corresponds to the vertical order $(a,b)=(0,1)$ (single jammer source using the STAP nomenclature), with some arbitrary modulation frequency $\omega$ and modulating process $s(m)$. In that case
\begin{eqnarray}
e^{(0,1)}(n, m)= s(m) \exp \biggl ( {j {\omega} n} \biggr )
\end{eqnarray}
The vector form representation of the finite sample of this evanescent field is
\begin{eqnarray}
{\bf e}^{(0,1)}&=&\left [
s(0),s(1),\dots,s(M-1), s(0) \exp  ( {j {\omega}}  ), \dots,s(M-1)\exp ( {j {\omega}} ) ,\dots \right. \nonumber \\
&&  \left.  s(0) \exp  ( {j {\omega}(N-1)} ) \dots,s(M-1)\exp ( {j {\omega}(N-1)} )\right ]^T.
\end{eqnarray}
Since the modulating frequency $\omega$ is a deterministic constant, it is obvious that ${\bf e}^{(0,1)}$ is comprised of only $M$ independent random variables. Therefore, the rank of $\lam^{(0,1)} = E \bigg [\ve^{(0,1)} \left (\ve^{(0,1)}\right)^H \bigg ]$ is also $M$.

{\it The aim of this paper is to derive an expression for the rank of the low-rank covariance matrix $\lam$ of the evanescent vector ${\bf e}$, in the general case (\ref{eq1a})}.

%%%%%%%%%%%%%%%%%%%%%%%%%%%%%%%%%%%%%%%%%%%%%%%%%%%%%%
%
%       Rank of Covariance
%
%%%%%%%%%%%%%%%%%%%%%%%%%%%%%%%%%%%%%%%%%%%%%%%%%%%%%%

\section{The Rank of the Covariance Matrix of an Evanescent Field}
\label{comp_rank}
 In this section we  derive an expression for  the rank of  the
covariance matrix $\lam$. In order to do so, we have to find and quantify the linear dependencies between the samples of ${\bf e}$. Unfortunately, for arbitrary spectral support parameters and multiple evanescent components, this task involves tedious calculations. The
results of the entire analysis in this section can be summarized by the
following theorem:

\begin{theorem} Let $\ve$ be a vector-form representation of a  finite sample from a sum of evanescent random fields, given by (\ref{eq1a})-(\ref{c16-1}). Then, the rank of its covariance matrix, $\lam$, is given by
\begin{equation}
rank (\lam)=\min\left(NM, \biggl[
N\sum_{q=1}^Q |a_q|+M\sum_{q=1}^Q |b_q|-\sum_{q=1}^Q |a_q|\sum_{q=1}^Q |b_q|\biggr]\right).
\label{th1a}
\end{equation}
\end{theorem}

Even though the evaluations in the next subsections are  technical in
nature, the obtained result is surprisingly interesting. Hence, before addressing the proof itself let us make some comments. From
Theorem 1, it is clear that the rank of the covariance matrix of a
finite sample from an evanescent random field is completely
determined by the spectral support parameters $(a_q,b_q)$ of the
different evanescent components, while it is
independent of the other parameters of the evanescent fields, such
as the parameters of the modulating processes, $\{s_q \}$,
or the modulation frequencies, $\omega_q$.

Moreover, one can easily observe that the well-known Brennan rule for the rank of the low-rank clutter covariance matrix in the STAP framework, \cite{ward} as well as the rank of the covariance matrix of the jamming signals are special cases of this theorem.
The Brennan rule states that the rank of the clutter covariance $\lam_{clut}$ is given by:
\begin{equation}
rank (\lam_{clut})=\lfloor N+M\beta - \beta\rfloor
\end{equation}
where $\beta$ is the  slope of the clutter ridge orientation in the angle-Doppler domain, and $\lfloor \rfloor$  denotes rounding to the nearest integer. It is easy to see that this formula is a  special case of the above theorem when only a single evanescent field is observed, and its spectral support parameters are $(a,b)=(1,\beta)$.  The rank of the jamming covariance matrix is,  \cite{ward}:
\begin{equation}
rank (\lam_{jam})=MJ
\end{equation}
where $J$ is a number of sources. Since the spectral support of a single jammer in the angle-Doppler domain is a line parallel to the Doppler axis, and since all jammers are mutually orthogonal, they can be modeled as  $J$  vertical evanescent components with spectral support parameters $(a,b)=(0,1)$, such that the rank of the covariance matrix of each individual jammer is $M$ as in the above example.

\subsection{Rank derivations}
In this subsection we  prove Theorem 1. The derivation provides an insight into the structure of the covariance matrix, and  explains the nature of its low-rank. Moreover, we explicitly show how columns of the covariance matrix that can be represented as linear combinations of other columns  are formed, which yields  its  low-rank.
%Nevertheless, the complete proof is very tedious. We refer the reader to extended version of this paper where complete stage-wise proof of the theorem is presented (ArXiv).

Rewriting (\ref{eq1a}) in a vector form we have ${\bf e}=\sum_{q=1}^{Q}\ve_q$, where
\begin{eqnarray}
{\bf e}_q&=&\left [
e_q(0,0),\dots,e_q(0,M-1),e_q(1,0),\dots,e_q(1,M-1), \right. \nonumber \\
&&  \left.  \dots, e_q(N-1, 0) , \dots, e_q(N-1,M-1)
\right ]^T . \label{c16}
\end{eqnarray}

Let
\begin{eqnarray}
&& \hspace{-.5in}\bxi_q=\left [
s_q(0),s_q(b_q),\dots,s_q((M-1)b_q), s_q(a_q),s_q(a_q+b_q),\dots,s_q(a_q+(M-1)b_q), \right. \nonumber \\
&& \hspace{-.2in} \left.
\dots,s_q((N-1)a_q),s_q((N-1)a_q+b_q),\dots,s_q((N-1)a_q+(M-1)b_q)
 \right ]^T  \label{c16-1a}
\end{eqnarray}
be the vector whose elements are the observed samples from the 1-D
modulating process $\{s_q\}$. Define
\begin{eqnarray}
\vv_q&=& \left [ \ 0, \ d_q,\dots,
\ (M-1)d_q,\ c_q,\ c_q+d_q,\dots,
\ c_q+(M-1)d_q ,\dots, \right. \nonumber \\
&&  \left. (N-1)c_q,\ (N-1)c_q+d_q,\ \dots, \ (N-1)c_q+(M-1)d_q\  \right]^T. \label{c16-2c}
\end{eqnarray}
Let
\begin{equation}
\vD_q=diag\biggl(\exp ( - j {\omega_q}\vv_q )\biggr).
\end{equation}
be an $NM\times NM$ diagonal matrix.
 Thus, using (\ref{eq1b}), we have that
\begin{equation}
{\bf e}_q=\vD_q^H \bxi_q
 \ . \label{c16-3}
\end{equation}

Let $\vs_q$ be a $(N-1)|a_q|+(M-1)|b_q| +1$ dimensional column
vector of consecutive samples of the 1-D modulating process
$\{s_q \}$. For the case in which $a_q >0$ and $b_q < 0$,
$\vs_q$ is defined as
\begin{eqnarray}
\vs_q&=&\left [ s_q((M-1)b_q), \dots, \dots,
s_q((N-1)a_q)
 \right ]^T \ ,
\label{c16-4}
\end{eqnarray}
while for the case in which $a_q \ge 0$ and $b_q \ge 0$, $\vs_q$
is defined as
\begin{eqnarray}
\vs_q&=&\left [ s_q(0), \dots, \dots,
s_q((N-1)a_q+(M-1)b_q)
 \right ]^T \ .
\label{c16-5}
\end{eqnarray}
Thus for any $(a_q,b_q)$ we have that
\begin{equation}
\bxi_q= \vA_q^T \vs_q  \label{c16-6a}
\end{equation}
and
\begin{equation}
{\bf e}_q= \vD_q^H\vA_q^T \vs_q,
\label{c16-6}
\end{equation}
where $\vA_q$ is a real-valued $[(N-1)|a_q|+(M-1)|b_q|+1] \times NM$
rectangular matrix where each of its columns has a single element whose value is ``1", while all the others are zero.
%Every column of $\vA_q$ corresponds to an element in the vector $\bxi_q$ and consequently in the vector $\ve$, while every row corresponds to an element of $\vs_q$.
Thus, each column of $\vA_q$ ``chooses" the single element from the vector $\vs_q$ that contributes to the corresponding
element of the vector $\bxi_q$ . Due to boundary effects,
resulting from the finiteness of the observation, not {\it all} of
the elements of the vector $\vs_q $ contribute to the vector $\bxi_q$, unless $|a_q|\le 1$ or $|b_q|\le
1$.
%In other words, for some arbitrary $a$ and $b$ there are missing
%samples from the modulating process $\{s_q \}$ in $\bxi_q$.
Hence some rows of the matrix $\vA_q$ may
contain only zeros. On the other hand, whenever $n a_q+m b_q=k a_q+\ell b_q$
for some integers $n,m,k,\ell$ such that $0 \le n, k \le N-1$ and $0
\le m, \ell \le M-1$, the same sample from the modulating process
$\{s_q \}$ is duplicated in the elements of
$\bxi_q$. Therefore, the number of {\it distinct} columns in
$\vA_q$ is equal to the number of elements of $\vs_q$
that appear in $\bxi_q$, \ie, the number of {\it distinct}
samples from the random process $\{s_q \}$ that are found in
an observed finite sample of an evanescent field of dimensions $N\times M$. The matrix
$\vA_q$ depends only on $(a_q,b_q)$ and is independent of the modulation frequency $\omega_q$ or the modulating process $\{s_q \}$.

The rank of covariance matrix $\lam$ is strongly related to the number of {\it distinct} samples from the random processes $\{s_q \}$ for all $1\leq q \leq Q$ which can be found in the evanescent vector $\ve$. Therefore, the rank of $\lam$ is tightly related to the ranks of the matrices $\vA_q$, $1\leq q \leq Q$.

Let ${\vR_q}$ be the covariance matrix of the vector $\vs_q$ \ie,
\begin{equation}
{\vR_q}= E \bigg [\vs_q \left (\vs_q\right
)^H \bigg ] \label{c16-7}.
\end{equation}
The matrix $\vR_q$ is full rank positive definite since the process
$\{s_q\}$ is purely-indeterministic.
% and hence admits
%innovation driven infinite order moving average representation.
Since the evanescent components $\{e_q\}$ are mutually
orthogonal we conclude that $\lam$, the covariance matrix of $\ve $,
 has the form
\begin{equation}
\lam= E \bigg [\ve \left (\ve\right )^H \bigg ]=
\sum_{q=1}^{Q} \lam_q \ , \label{e3}
\end{equation}
where $\lam_q$ is the covariance matrix of $ \ve_q$.
Using (\ref{c16-6}) and  (\ref{c16-7}) we find that
\begin{eqnarray}
\lam_q &=& E \bigg [\ve_q \left (\ve_q\right
)^H \bigg ]=\vD_q^H\vA_q^T\vR_q \vA_q\vD_q \ \ . \label{e4a}
\end{eqnarray}
Finally,
\begin{equation}
\lam= \sum_{q=1}^{Q} \vD_q^H\vA_q^T\vR_q \vA_q\vD_q \ .
\end{equation}
One can rewrite the above expression in a block-matrix
form
\begin{equation}
\lam=\vC_Q^H\vR\vC_Q
\end{equation}
where
\begin{equation}
\vC_Q = \bigl [\vD_1^H\vA_1^T \hdots \vD_Q^H\vA_Q^T \bigr ]^H,
\end{equation}
and
\begin{equation}
\vR =diag \bigl(\bigl [\vR_1 \dots  \vR_Q\bigr ] \bigr ).
\end{equation}
is a block-diagonal matrix with the matrices $\vR_q$, $1\leq q \leq Q$ on its diagonal, and zeros elsewhere.

Since the covariance matrices $\vR_q$ are full rank positive definite, the block-diagonal
matrix $\vR$ is full rank positive
definite as well. Hence by observation 7.1.6 \cite{horn} we have
\begin{equation}
rank(\lam)= rank
\left(\vC_Q\right). \label{end2b}
\end{equation}

The matrix $\vC_Q$ has exactly $NM$ columns, such that each one of its columns corresponds to an entry in the evanescent vector $\ve$, or similarly, each one of its columns corresponds to a point on the original $N \times M$ lattice $D = \{(n,m)\in
\mathbb{Z}^2: 0 \le n \le N - 1,\,0 \le m \le M - 1\} $. More specifically, the $n(M-1)+m$ column of $\vC_Q$ corresponds to the $n(M-1)+m$ element of $\ve$ which represents the evanescent field sample at the $(n,m)$  lattice point. (Note that we enumerate the columns starting from zero). {\it In the following we will adopt the abbreviation {\bf $[n,m]$} for indexing the $n(M-1)+m$  column of  a matrix}.

To gain more understanding on the  structure of $\vC_Q$ let us examine the different matrices $\vC_Q$ is comprised of. We begin with $\vA_q$ for some $1\leq q \leq Q$:
By construction (see (\ref{c16-6}) and the following explanation) all columns of
$\vA_q$ are unit vectors, where the single ``1" entry in
each column chooses the single element from the vector
$\vs_q$ that contributes to $e(n,m)$ - the evanescent
field sample at
$(n,m)$ . The single
non-zero entry in the $[n,m]$ column of $\vA_q$ is located in the
$k$'th row where $na_q+mb_q=k$ (we allow negative indexed rows in
the case where $b_q<0$). For example if $a_q>0$ and $b_q>0$, the
matrix $\vA_q$ is given by
\begin{equation}
\hspace{-.7in}\begin{array}{c}
          \\
          0\\
          \\
        k \\
         \\
 (N-1)a_q+(M-1)b_q
      \end{array}
\begin{array}{c}
\begin{array}{ccccc}
 [0,0] & \cdots & [n,m] &  \cdots & [N-1,M-1]
\end{array}
\\
\begin{bmatrix}
  1 &  \cdots     & \cdots & 0 &   \cdots&  \cdots&  \cdots& 0 &  \\
  \vdots &       &  &  \vdots&  &  &    &  \vdots&  \\
  \vdots &  &  & 1 &  &  &    &  \vdots&  \\
  \vdots &  &  &  \vdots&  &  &    &  \vdots&  \\
  0 &    \cdots   &  \cdots&  0&  \cdots&  \cdots& \cdots   &  1&  \\
\end{bmatrix}
\\
\end{array}
\end{equation}

Let $(n^*,m^*)$ be a solution to the {\it linear Diophantine equation} $na_q+mb_q=k$. Then,
the equation is also satisfied by $n=n^*+tb_q$ and $m=m^*-ta_q$,
where $t$ is an arbitrary integer. Since $(a_q,b_q)$ are coprime
integers these are the only possible solutions. It means that as soon as
$(n+tb_q,m-ta_q) \in D$, the corresponding $[n+tb_q,m-ta_q]$ column of $\vA_q$ will be equal to its $[n,m]$ column. To find the rank of $\vA_q$ we have to evaluate the number of linearly independent columns, \ie, the number of distinct elements of $s_q(na_q+mb_q)$  $(n,m) \in D$ which contribute to $\ve$.

Since $\vD_q$ is a diagonal matrix, the structure of $\vA_q\vD_q$ is similar to the structure of $\vA_q$ with the only difference being that instead ``1" in each column, we have the appropriate exponential coefficient. Therefore, each column of the matrix $\vC_Q$ has exactly $Q$ non-zero elements.

Next, let us concatenate the matrices $\vA_p$ and $\vA_q$, where $1\leq p\neq q \leq Q$ and examine the structure of resulting matrix \begin{equation}
\tilde \vC_{pq}=[\vA_p^T\ \vA_q^T\ ]^T.
\end{equation}
%\begin{equation}
%\tilde \vC_{pq}=\begin{bmatrix}
%               \vA_p\\
%               \vA_q\\
%             \end{bmatrix}.
%\end{equation}
As before, let us consider the structure of some arbitrary  $[n,m]$ column of this matrix. It has two non-zero entries: On the $k_p$ row of $\vA_p$ and on the $k_q$ row of $ \vA_q$, where $(n,m)$ satisfies
\begin{equation}
   na_p+mb_p=k_p,
   \label{k1}
 \end{equation}
and
\begin{equation}
 na_q+mb_q = k_q.
 \label{k2}
 \end{equation}

Next, we note that the pair $(n+tb_p,m-ta_p)$ satisfies the
linear Diophantine equation (\ref{k1}) for any integer $t$. Therefore, for $t_p$ such that $(n+t_pb_p,m-t_pa_p) \in D$, the $[n+t_pb_p,m-t_pa_p]$ column of $\vA_p$ has a ``1" entry, at the same row as the $[n,m]$ column. However, $(n+t_pb_p,m-t_pa_p)$ also satisfies the linear
 Diophantine
equation
\begin{equation}
   (n+t_pb_p)a_q+(m-t_pa_p)b_q= \ell_q.
   \label{l2}
\end{equation}
Hence, the $[n+t_pb_p,m-t_pa_p]$ column of $ \vA_q$ has a ``1" entry on its  $\ell_q$ row.

Similarly, since $(n+tb_q,m-ta_q)$ satisfies the
linear Diophantine equation (\ref{k2}) for any integer $t$, for $t_q$ such that $(n+t_qb_q,m-t_qa_q) \in D$ we have that the $[n+t_qb_q,m-t_qa_q]$ column of $\vA_q$ has ``1" at the same row as the $[n,m]$ column. Since,
\begin{equation}
   (n+t_qb_q)a_p+(m-t_qa_q)b_p= \ell_p,
   \label{l1}
 \end{equation}
the $[n+t_qb_q,m-t_qa_q]$ column of $\vA_p$ has ``1" on its  $\ell_p$ row.
Moreover, one can observe that for a pair of integers $(t_p, t_q)$ such that $(n+t_pb_p+t_qb_q,m-t_pa_p-t_qa_q) \in D $, the pair $(n+t_pb_p+t_qb_q,m-t_pa_p-t_qa_q)$ simultaneously satisfies (\ref{l2}) and
(\ref{l1}):
\begin{eqnarray}
   (n+t_pb_p+t_qb_q)a_p+(m-t_pa_p-t_qa_q)b_p &=& \ell_p \nonumber \\
   (n+t_pb_p+t_qb_q)a_q+(m-t_pa_p-t_qa_q)b_q &=& \ell_q
   \label{eq34_e}
 \end{eqnarray}
Therefore the $[n+t_pb_p+t_qb_q,m-t_pa_p-t_qa_q]$ column of of $\vA_p$ has ``1" on its $\ell_p$ row, and the same column of $\vA_q$ has ``1" on its $\ell_q$ row.

Finally,  we can represent the $[n,m]$ column of $\tilde \vC_{pq}$ by a linear combination of its other columns:
 \begin{eqnarray}
[n,m]&=&[n+t_pb_p,m-t_pa_p]+[n+t_qb_q,m-t_qa_q]-\nonumber\\
&&[n+t_pb_p+t_qb_q,m-t_pa_p-t_qa_q],
\end{eqnarray}
or in a more detailed form by
\begin{equation}
\begin{array}{c}
          \\
          k_p\\
          \\
         k_q \\
           \\
           \\
       \end{array}
       \left[
  \begin{array}{c}
     \\
     1\\
     \\
     1\\
     \\
     \\
  \end{array}
\right]=\begin{array}{c}
          \\
         k_p\\
          \\
          \\
           \ell_q\\
           \\
       \end{array}
       \left[
  \begin{array}{c}
     \\
     1\\
     \\
     \\
     1\\
     \\
  \end{array}
\right]+\begin{array}{c}
          \\
          \\
          \ell_p\\
          k_q \\
           \\
           \\
       \end{array}
       \left[
  \begin{array}{c}
     \\
     \\
    1 \\
     1\\
     \\
     \\
  \end{array}
\right]-\begin{array}{c}
          \\
          \\
          \ell_p\\
           \\
           \ell_q\\
           \\
       \end{array}
       \left[
  \begin{array}{c}
     \\
     \\
    1 \\
     \\
     1\\
     \\
  \end{array}
\right]
\end{equation}

Let $T_{pq}^{(n,m)}$ be the set of {\it all} the integer pairs $(t_p,t_q)$
such that
$(n+t_pb_p,m-t_pa_p),(n+t_qb_q,m-t_qa_q),(n+t_pb_p+t_qb_q,m-t_pa_p-t_qa_q)
\in D$. Clearly, the set $T_{pq}^{(n,m)}$ is non-empty since $(0,0)\in
T_{pq}^{(n,m)}$, and it corresponds to a trivial representation of the $[n,m]$ column by itself.
If $|T_{pq}^{(n,m)}|>1$ then  the $[n,m]$ column has non-trivial linear representation by other columns.

Recall however, that the matrix $\vC_Q$ is comprised of blocks where each block is of the form  $\vA_q\vD_q$. Consider next the concatenation of two such blocks $\vA_p\vD_p$ and $\vA_q\vD_q$,
\begin{equation}
\vC_{pq}=[\vD_p^H\vA_p^T\ \vD_q^H\vA_q^T\ ]^H.
\label{cq}
\end{equation}
Keeping in mind that by definition $a_pd_p-b_pc_p=1$ and $a_qd_q-b_qc_q=1$, it is easy to check that the replacement of the ``1" in the columns of $\tilde \vC_{pq}$ by  exponentials as in (\ref{cq}) will
only affect the coefficients of the linear combination. Indeed, the linear combination of columns in this case has the form
\begin{eqnarray}
[n,m]&=&[n+t_pb_p,m-t_pa_p]\exp(  {j
\omega_pt_p}) +[n+t_qb_q,m-t_qa_q]\exp(  {j
\omega_qt_q})\nonumber
\\
&-&[n+t_pb_p+t_qb_q,m-t_pa_p-t_qa_q]\exp( {j
[\omega_pt_p +\omega_qt_q]}).
\label{new_eq0}
\end{eqnarray}
One may also notice that if $(a_p,b_p)=(a_q,b_q)$ we have that $k_p=k_q=\ell_p=\ell_q$. However, since in this case $\omega_p\neq\omega_q$ the linear combination in (\ref{new_eq0}) is still valid and non-trivial.

It is clear that the linear dependencies of columns of $\vC_{Q}$  %which correspond to a larger $Q$
%will further complicate the linear dependencies of its columns, but these dependencies will still be
are governed by the same simple laws:
Let $T_Q^{(n,m)}$ be a set of all $Q$-tuples of integers
$(t_1,\ldots,t_Q)$ defined as follows: For any $1\leq q \leq Q$, let
$(i_1,\ldots,i_q)$ be a set of $q$ indices, such that $1\leq i_k
\leq Q$ for all $1\leq k \leq q$, and
$$(n+t_{i_1}b_{i_1}+\ldots+
t_{i_q}b_{i_q},m-t_{i_1}a_{i_1}-\ldots-t_{i_q}a_{i_q})\in D.$$
Clearly, the set $T_Q^{(n,m)}$ is non-empty since $(0,\ldots,0)\in
T_Q^{(n,m)}$.
Let $(n,m)\in D$ be an arbitrary lattice point and let
$[n,m]$ be its corresponding column in $\vC_Q$. Then, the $[n,m]$
column can be represented by the linear combination
\begin{eqnarray}
&&\hspace{-.3in}[n,m]= \nonumber \\
&&\hspace{-.3in}\sum_{q=1}^Q(-1)^{q-1}
\mathop{\underbrace{\sum_{{i_1}=1}^{Q-q+1}\ldots\sum_{{i_q}={i_{q-1}}+1}^{Q}}}\limits_{\mbox{\emph{q}
sums}}
[n+t_{i_1}b_{i_1}+\ldots+t_{i_q}b_{i_q},m-t_{i_1}a_{i_1}-\ldots-t_{i_q}a_{i_q}]\cdot
\nonumber \\
&&\hspace{1.6in}\cdot \exp\left( {j
{[\omega_{i_1}t_{i_1}+\ldots+\omega_{i_q}t_{i_q}]}}\right),
\label{prop2a}
\end{eqnarray}
where $(t_1,\ldots,t_Q)\in T_Q^{(n,m)}$. The details of this derivation are presented in Appendix A.

\begin{figure}
  \begin{center}
  \includegraphics[width=13cm]{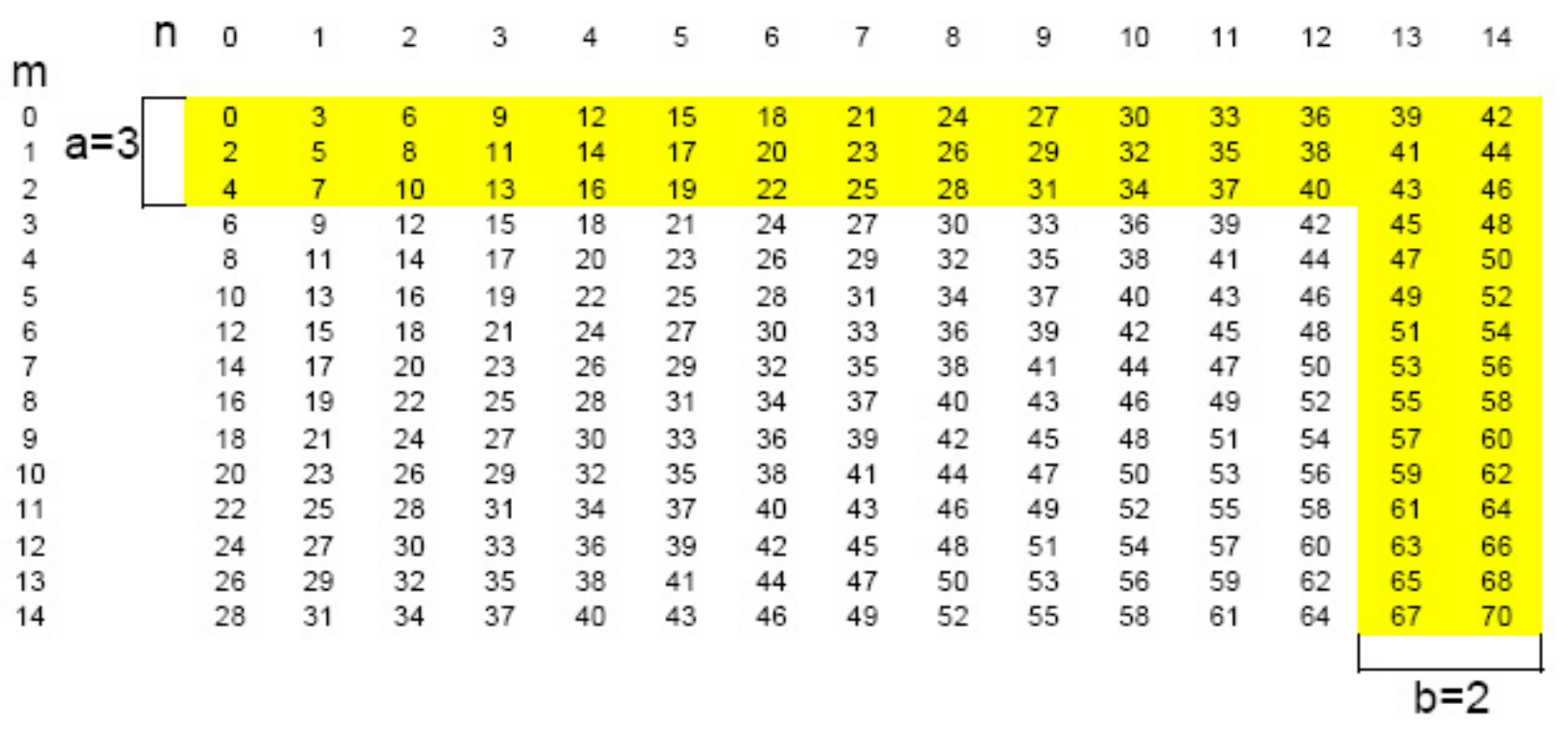}\\
  \centerline{(a)}
  \vspace{0.2in}
  \includegraphics[width=13cm]{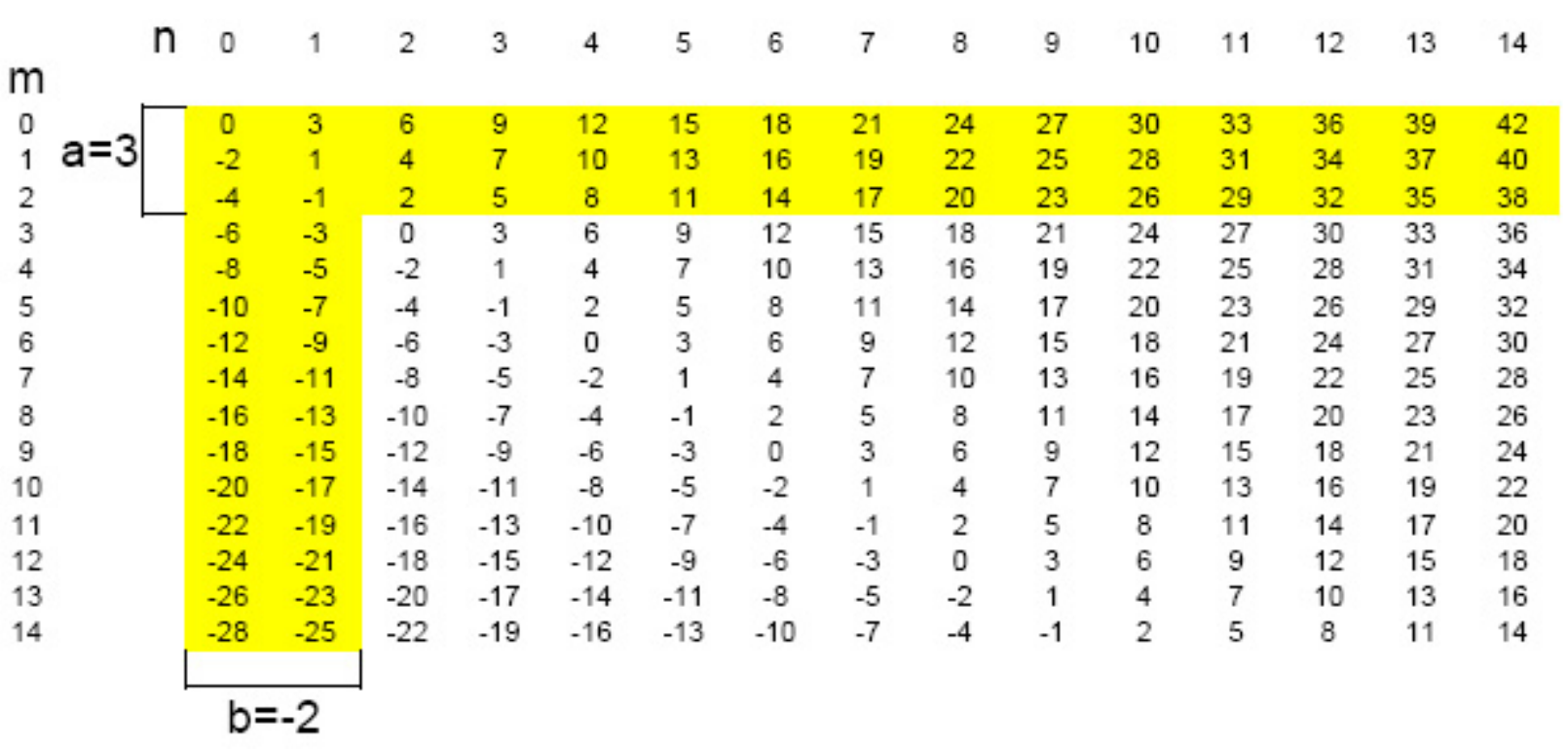}\\
  \centerline{(b)}
  \caption{$N=M=15, a=3, b=\pm 2$:  (a) The indices $k=na+mb$ of the observation on
  $\{s^{(3,2)}(k) \}$; (b) The indices $k=na+mb$ of the observation on $\{s^{(3,-2)}(k) \}$.
  The sets of distinct indices of $\{s^{(3,2)}(k) \}$ and $\{s^{(3,-2)}(k) \}$ are marked
in yellow. Every other sample in the field is identical to some
sample in the yellow area.}\label{a}
 \end{center}
\end{figure}

Following the foregoing analysis of the linear dependencies between the columns of $\vC_{Q}$, we next count its linearly independent columns in order to derive the rank of $\vC_{Q}$.  Let us first count the number of independent columns of $\vA_q\vD_q$. As  mentioned earlier, this number is equal the number of distinct samples from $s_q(na_q+mb_q)$,  $(n,m) \in D$ that contribute to $\ve$. In other words, this is the number of different indices $k$, such that $na_q+mb_q=k$ where $(n,m) \in D$, and it can be easily calculated based on the dimensions of  $D$ (see  Figure \ref{a} for an illustrative
example). Indeed, a new sample from the random process $\{s_q\}$ may be introduced only on the first $a_q$ rows (since
$a_q\geq 0$) and the last (first) $|b_q|$ columns (last if $b_q\geq
0$ and first if $b_q< 0$) of the observed finite dimensional field, while the rest of
the field is filled by replicas of these samples. We thus count
$Na_q$ distinct samples in the first $a_q$ rows and $M|b_q|$ distinct
samples in the first (last) $|b_q|$ columns. However, on the intersection of
these rows and columns $|a_qb_q|$ samples are counted twice.
Finally, the total number of {\it distinct} samples from the random
process $\{s_q \}$ that are found in an observed field of
dimensions $N\times M$ (which is equal to the rank of $\vA_q$ and the rank of $\vA_q\vD_q$) is
given by
\begin{equation}
r_q=Na_q+M|b_q|-|a_qb_q|.
\label{e701}
\end{equation}

Similarly, it can be shown that the number of linearly independent columns of $\vA_p\vD_p$ is $Na_p+M|b_p|-|a_pb_p|$. Let us next count the number of linearly independent columns of $\vC_{pq}$.

\begin{figure}
  \begin{center}
  \includegraphics[width=13cm]{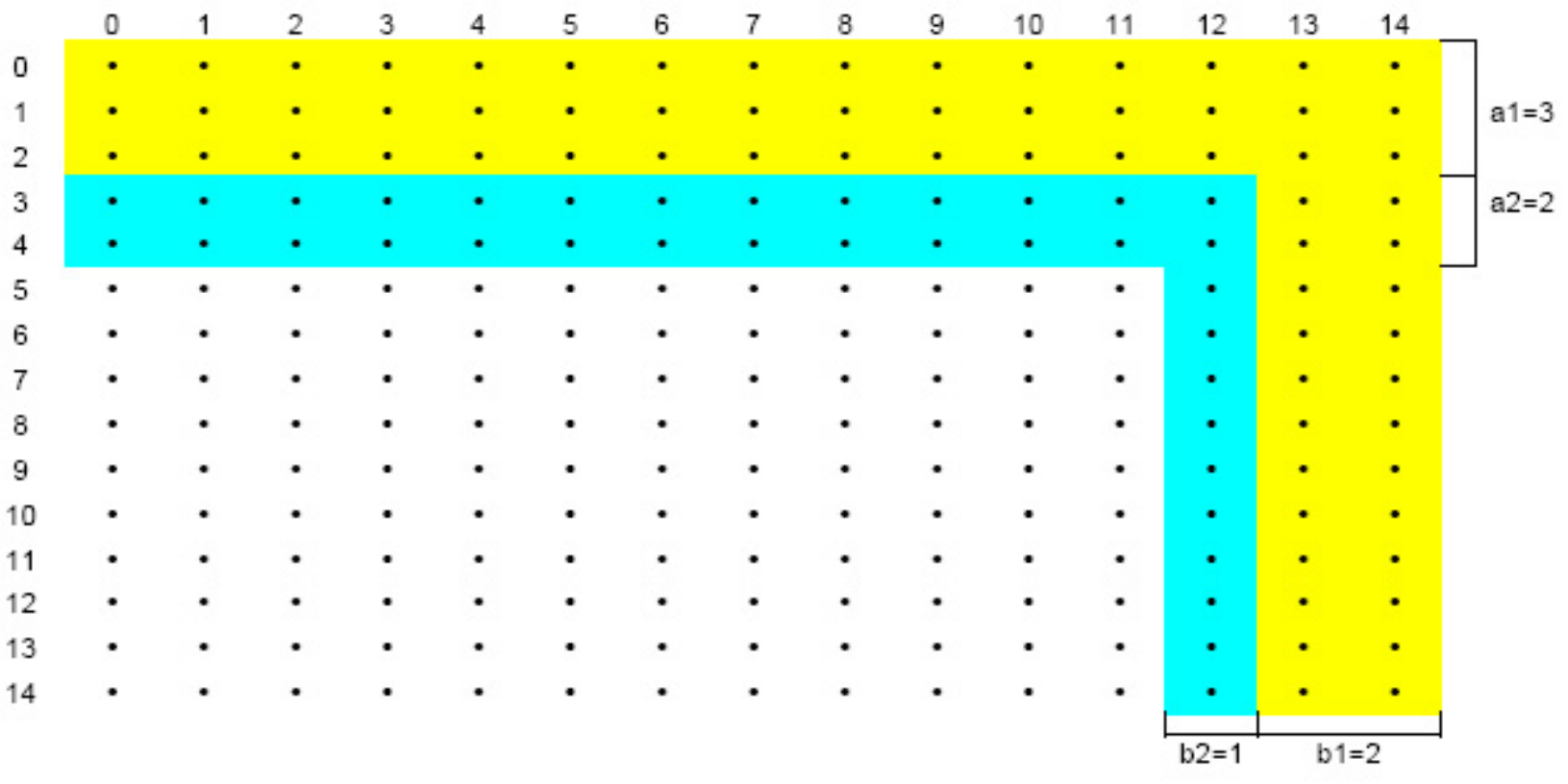}\\
  \centerline{(a)}
  \vspace{0.2in}
   \includegraphics[width=13cm]{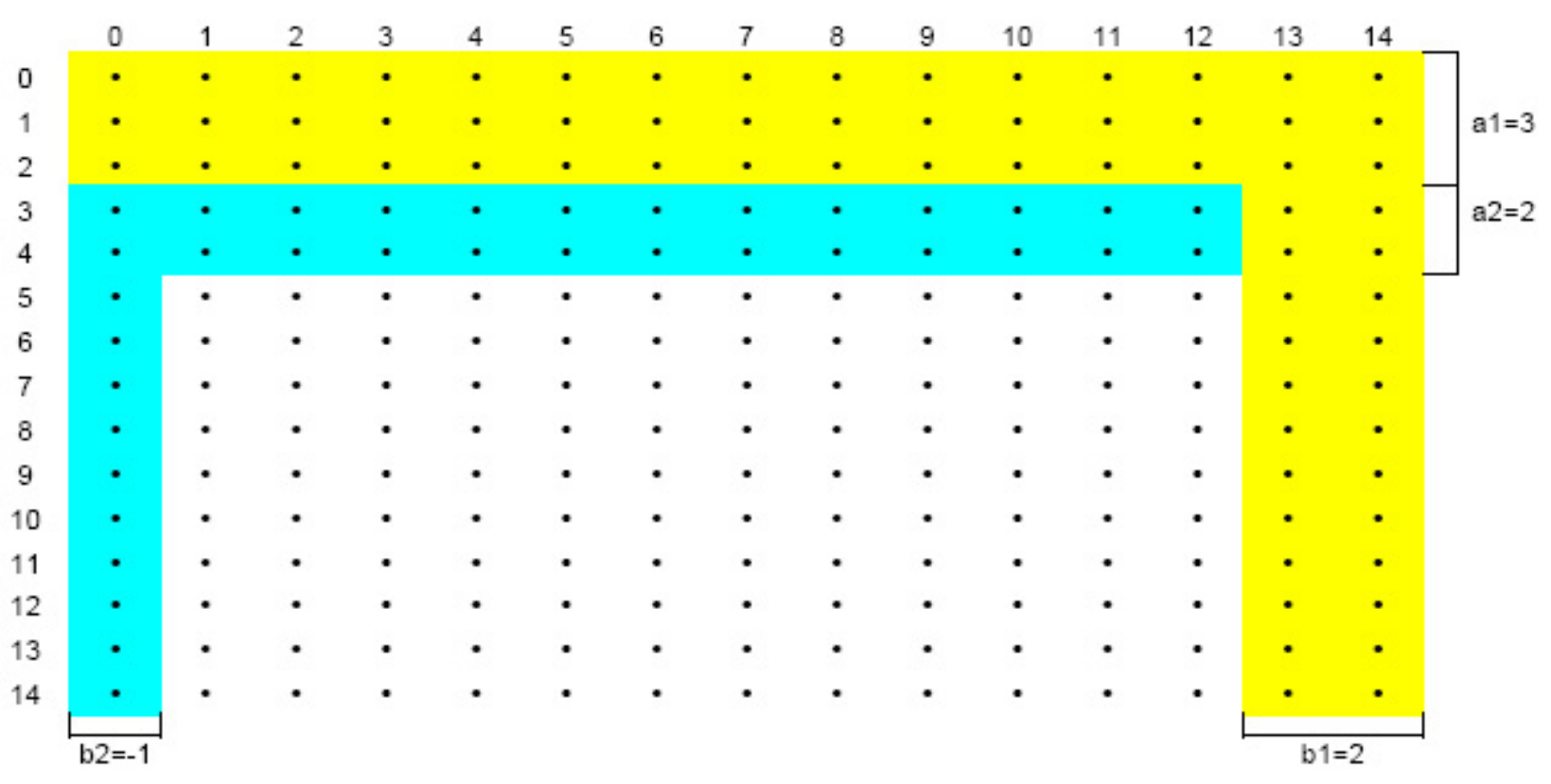}
   \centerline{(b)}
  \end{center}
  \caption{$N=M=15, a_1=3, b_1=2, a_2=2, b_2=\pm 1$. (a): The set of distinct samples of $s^{(3,2)}(n,m)$
  (in yellow) and $s^{(2,1)}(n,m)$ (in blue). (b): The set of distinct samples of $s^{(3,2)}(n,m)$
  (in yellow) and $s^{(2,-1)}(n,m)$ (in blue). In both cases, every other sample in the
  field is a linear combination of samples in the colored areas.} \label{b}
\end{figure}

Since $r_p$ columns of $\vA_p\vD_p$ are
linearly independent, the same columns of the concatenated matrix $\vC_{pq}$ are linearly independent as well. The remaining $NM-r_p=(N-|b_p|)(M-|a_p|)$ columns
may be considered to correspond to an $(N-|b_p|) \times (M-|a_p|)$
rectangular sub-lattice $D_1 = \{(n,m)\in \mathbb{Z}^2: 0 \le n \le N -
1-|b_p|,\,|a_p| \le m \le M - 1\} $ (see Figure \ref{b} as an
example), which is a subset of the original rectangular lattice (or
similarly, one can define  $D_1 = \{(n,m)\in \mathbb{Z}^2: |b_p| \le
n \le N - 1,\,|a_p| \le m \le M - 1\} $ which doesn't change the
reasoning of our arguments and only depends on a sign of $b_p$).
Repeating the same arguments as  those made above, one can show that the number of
{\it distinct} samples from the random process $\{s_q \}$
that are found in a sub-lattice $D_1$ is
\begin{equation}
\tilde{r_q}=(N-|b_p|)|a_q|+(M-|a_p|)|b_q|-|a_qb_q| \
. \label{y}
\end{equation}
This is the number of linearly independent columns which can be found in $\vC_{pq}$ in addition to the first $r_p$ columns.

 Let $D_2$ be the set of
$NM-r_p-\tilde r_q=(N-|b_p|-|b_p|)(M-|a_q|-|a_q|)$ lattice points
that remain after the removal from $D$ of the $r_p+\tilde r_q$
points corresponding to the  linearly independent columns of $\vC_{pq}$ (for simplicity and without limiting of the generality
of the results, we will discuss the case where $b_p>0$ and $b_q>0$,
as illustrated in Figure 2a (uncolored area)). Thus, $D_2 =
\{(n,m)\in \mathbb{Z}^2: 0 \le n \le N - 1-|b_p|-|b_q|,\,|a_p|+|a_q|
\le m \le M - 1\} $. It thus remains to be shown that all columns representing points in $D_2$ can be represented by a linear
combination of columns that correspond to points in  $D\setminus D_2$.

 Since the ``width" of
$D\setminus D_2$ is $|b_p|+|b_q|$ along the $n$-axis and
$|a_p|+|a_q|$ along the $m$-axis  (colored areas in Figure 4), for every $(n,m) \in D_2$ we have $(n + b_p,m - a_p),(n+ b_q,m - a_q),(n + b_p+b_q,m - a_p- a_q)\in D$. Thus, $(t_p,t_q)=(1,1) \in T_{pq}^{(n,m)}$, and as we have shown above, we can represent
$[n,m]$ by the linear combination
\begin{eqnarray}
[n,m]&=&[n+b_p,m-a_p]\exp(  {j
\omega_p}) +[n+b_q,m-a_q]\exp(  {j
\omega_q})\nonumber \\
&-&[n+b_p+b_q,m-a_p-a_q]\exp( {j
[\omega_p +\omega_q]}).
\end{eqnarray}

Continuing this construction recursively, it is obvious that for each point $(n,m) \in D_2$ we can find a pair $(t_p,t_q)\in T_{pq}^{(n,m)}$, and $(t_p,t_q) \neq (0,0)$ such that  $(n + t_pb_p,m - t_pa_p),(n+ t_qb_q,m
- t_qa_q),(n + t_pb_p+t_qb_q,m - t_pa_p- t_qa_q)\in D \setminus D_2$. On the other hand, for every point $(n,m) \in D \setminus D_2$ one can show that  $T_{pq}^{(n,m)}=\{(0,0)\}$, \ie, only the trivial linear combination exists. In other words, all the random variables indexed on $D \setminus D_2 $ correspond to linearly independent columns. Therefore, the number of linearly independent columns in $\vC_{pq}$ is
\begin{eqnarray}
&&rank(C_{pq})=| D \setminus D_2|=r_p+\tilde{r_q}\nonumber
\\
&&=N(|a_p|+|a_q|)+M(|b_p|+|b_q|)-(|a_p|+|a_q|)(|b_p|+|b_q|).
\label{ra}
\end{eqnarray}

\begin{figure}
  \begin{center}
  \includegraphics[width=13cm]{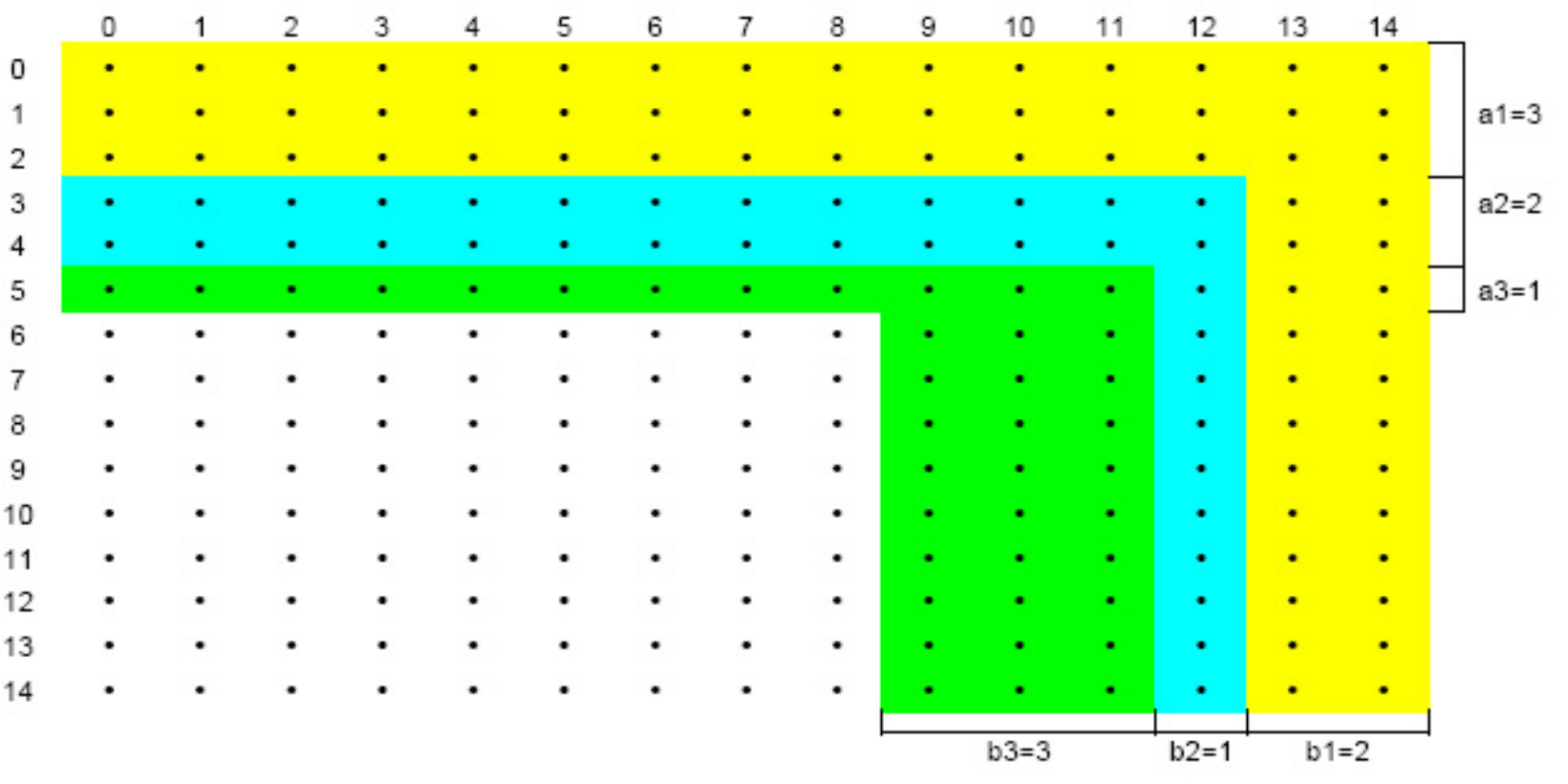}\\
  \end{center}
  \caption{Sets of distinct samples in the case of three evanescent components. $N=M=15,a_1=3, b_1=2, a_2=2, b_2=1,a_3=1,b_3=3$}\label{d3}
\end{figure}

The construction described above can be easily extended to the general case where we concatenate {\it all} the matrices which $\vC_Q$ is comprised of. See Figure \ref{d3} for
an example of a three component case. If one chooses a subset of the original lattice,
$$D_Q=\{(n,m)\in \mathbb{Z}^2: 0 \le n \le N -
1-\sum_{q=1}^Q|b_q|,\,\sum_{q=1}^Q|a_q| \le m \le M - 1\}, $$ which
remains after the removal of
\begin{equation}
N\sum_{q=1}^Q|a_q|+M\sum_{q=1}^Q|b_q|-\sum_{q=1}^Q|a_q|\sum_{q=1}^Q|b_q|
\end{equation}
lattice points (similarly to $D_2$ which remains after the removal
of the $r_p+\tilde r_q$ lattice points corresponding to the
independent columns of $\vC_{pq}$), one can repeat the same considerations as above and show that columns of $\vC_Q$ that correspond to the lattice points in $D\setminus D_Q$ are the {\it
only} linearly independent columns of $\vC_Q$. Thus,
\begin{eqnarray}
\hspace{-.3in}rank(\vC_Q)=|D\setminus D_Q|=
N\sum_{q=1}^Q|a_q|+M\sum_{q=1}^Q|b_q|-\sum_{q=1}^Q|a_q|\sum_{q=1}^Q|b_q|.
\label{end2a}
\end{eqnarray}

Finally, Since the rank of $\lam$ cannot exceed $NM$ (the dimension
 of the covariance matrix), $NM$ is an upper bound on the rank of
 $\lam$. Combining this and (\ref{end2a}) the rank of $\lam$ is given by (\ref{th1a}), which completes the proof.

%%%%%%%%%%%%%%%%%%%%%%%%%%%%%%%%%%%%%%%%%%%%%%%%%%%%%%
%
%       Real Valued evanescent
%
%%%%%%%%%%%%%%%%%%%%%%%%%%%%%%%%%%%%%%%%%%%%%%%%%%%%%%

\section{The Case of a Real Valued Evanescent Field}
\label{real_rank}

 In the case where a real valued evanesced field is
considered, we have
\begin{eqnarray}
e_q(n, m) &=&
 s_q(n a_q+m b_q) \cos ( {
{\omega_q} (n c_q +m d_q)}) \nonumber \\
&+& t_q(n a_q+m b_q) \sin ( { {\omega_q} (n c_q
+m d_q)} ), \label{en_e30a}
\end{eqnarray}
where the 1-D purely-indeterministic processes $ \lbrace  s_q \rbrace $, $ \lbrace s_p \rbrace
$, $ \lbrace t_q \rbrace $,
%\linebreak
$ \lbrace  t_p \rbrace $ are mutually
orthogonal for all $1\leq p \ne q \leq Q$, and for all
$q$ the processes $ \lbrace  s_q \rbrace $ and $
\lbrace  t_q \rbrace $ have an identical
autocorrelation function. Let $\vt_q$ be defined similarly
to $\vs_q$ in (\ref{c16-5}). Using similar notations as in (\ref{c16-6}) we have
\begin{equation}
{\bf e}_q= {\cal R}(\vD_q^H)\vA_q^T
\vs_q+{\cal I}(\vD_q^H)\vA_q^T \vt_q,
\label{c16-6b}
\end{equation}
where ${\cal R}$ and ${\cal I}$ denote real and imaginary parts
respectively. Finally, since the processes $ \lbrace  s_q
\rbrace $ and $ \lbrace  t_q \rbrace $ are mutually
orthogonal  and have an identical autocorrelation function we find
that
\begin{eqnarray}
\lam_q = E \bigg [\ve_q \left (\ve_q\right
)^T \bigg ]
={\cal R}(\vD_q^H)\vA_q^T\vR_q \vA_q{\cal R}(\vD_q)+{\cal I}(\vD_q^H)\vA_q^T\vR_q \vA_q{\cal I}(\vD_q) \label{re-e4a}
\end{eqnarray}
where
\begin{equation}
\vR_q= E \bigg [\vs_q \left (\vs_q\right
)^T \bigg ]=E \bigg [\vt_q \left (\vt_q\right )^T
\bigg ] \label{c16-7a}
\end{equation}
is positive definite since $\{s_q\}$ and $\{t_q\}$
are purely-indeterministic. %and hence admits innovation driven
%infinite order moving average representation.

Similarly to the case of a complex valued evanescent field,
the covariance matrix $\lam$ is given by
\begin{equation}
\lam= \sum_{q=1}^{Q} \lam_q \ .
\label{e3a}
\end{equation}

The derivation of the rank of the covariance matrix (\ref{e3a}) follows exactly the same lines as in the previous section, and the next corollary is immediate:

\begin{corollary}
Let $\ve$ be a vector-form representation of a finite sample from a sum of real-valued evanescent random fields, given by (\ref{eq1a}),(\ref{en_e30a}) and (\ref{c16-1}). Then, the rank of its covariance matrix, $\lam$, is given by
\begin{equation}
rank (\lam)=\min\left(NM, \biggl[
N\sum_{q=1}^Q 2|a_q|+M\sum_{q=1}^Q 2|b_q|-\sum_{q=1}^Q 2|a_q|\sum_{q=1}^Q 2|b_q|\biggr]\right).
\label{th1re}
\end{equation}
\end{corollary}

%%%%%%%%%%%%%%%%%%%%%%%%%%%%%%%%%%%%%%%%%%%%%%%%%%%%%%
%
%       CONCLUSION
%
%%%%%%%%%%%%%%%%%%%%%%%%%%%%%%%%%%%%%%%%%%%%%%%%%%%%%%

\section{Conclusion}
\label{Concl}
 We have considered the problem of evaluating the rank
of the covariance matrix of a finite sample  from an evanescent
random field. We have analytically derived the rank formula and have
shown that the rank of the covariance matrix of this finite sample
from the evanescent random field is completely determined by the
evanescent field spectral support parameters and is independent of
all other parameters of the field. Thus, for example, the problem of evaluating
the rank of the low-rank covariance matrix of the interference in
 space time adaptive processing (STAP) of radar data may be solved
as a by-product of estimating only the spectral support parameters
of the interference components, when we employ a parametric model of the STAP data which is based on Wold decomposition, \cite{STAP}. Thus , this formula generalizes a well known result known as the Brennan rule
for the rank of the clutter covariance matrix in space-time adaptive
processing of airborne radar data. The derived rank formula may be employed in a wide
range of applications in radar signal processing as well as in other areas of signal and image processing.

\section{Appendix A}
To derive (\ref{prop2a}) let us choose an arbitrary lattice point $(n,m)\in D$,
and hence a corresponding column $[n,m]$. Exactly as in the two
component case, this column is associated with the random variable
$e(n,m)$. In fact we are looking for a representation of this random
variable by a linear combination of other random variables indexed
on $D$.

The first term in
the desired linear combination is a sum of $Q$ columns
\begin{eqnarray}
\sum_{i=1}^{Q}[n+t_ib_i,m-t_ia_i]\exp(j
\omega_it_i). \label{lc1}
\end{eqnarray}

This sum creates a new column vector. Similarly to the two component
case, this vector is composed of the $Q$ non-zero elements of the
$[n,m]$ column (similarly to the elements in rows $k_p$ and $k_q$ of
$\vC_{pq}$), but in addition it includes the {\it
undesired} elements (similar to the elements in rows $\ell_p$ and
$\ell_q$). The total number of contributed undesired elements is
$Q(Q-1)$. Each two pairs $(a_i,b_i)$ and $(a_j,b_j)$, $i \neq j$,
contribute a pair of undesired elements (one in $\vA_i\vD_i$ and one in
$\vA_j\vD_j$), which can be eliminated by subtraction of the
$[n+t_ib_i+t_jb_j,m-t_ia_i-t_ja_j]$ column multiplied by the
appropriate exponential coefficient, since
\begin{eqnarray}
&&\hspace{-.8in}(n+t_ib_i)a_j+(m-t_ia_i)b_j=(n+t_ib_i+t_jb_j)a_j
+(m-t_ia_i-t_ja_j)b_j \nonumber \\
&&\hspace{-.8in}(n+t_jb_j)a_i+(m-t_ja_j)b_i=(n+t_ib_i+t_jb_j)a_i
+(m-t_ia_i-t_ja_j)b_i
\end{eqnarray}
(See also (\ref{eq34_e}) for the equivalent scenario in the two
component case).

To eliminate all these undesired elements we subtract from the
vector in (\ref{lc1}) the sum of $\binom {Q}{2}$ such columns (half
the number of contributed undesired elements), and the result is
\begin{eqnarray}
 \label{lc2}
&&\hspace{-.5in} \sum_{i=1}^{Q}[n+t_ib_i,m-t_ia_i]\exp( {j
{\omega_it_i}}) -  \\
&&\sum_{i=1}^{Q-1}\sum_{j=i+1}^{Q}[n+t_ib_i+t_jb_j,m-t_ia_i-t_ja_j]\exp\left(
{j {[\omega_it_i+\omega_jt_j]}}\right)\nonumber
\end{eqnarray}
However, the resulting column now contains a new kind of {
undesired} elements. Substraction of the
$[n+t_ib_i+t_jb_j,m-t_ia_i-t_ja_j]$ column has eliminated $2$
undesired elements in $\vA_i\vD_i$ and $\vA_j\vD_j$, but at the same time has
created a new undesired element in every $\vA_k\vD_k$, such that $i\neq j
\neq k$. A total of $Q-2$ new undesired elements have been created.
Clearly, these elements are negative and their total number is
$\binom {Q}{2}(Q-2)=3\binom {Q}{3}$. To eliminate the contribution
of these elements we add the
$[n+t_ib_i+t_jb_j+t_kb_k,m-t_ia_i-t_ja_j-t_ka_k]$ column with an
appropriate exponential coefficient which eliminates the undesired
element from $\vA_k\vD_k$. At the same time this action is also canceling
the undesired result of subtracting
$[n+t_ib_i+t_kb_k,m-t_ia_i-t_ka_k]$ that appears in $\vA_j\vD_j$, and the
undesired result of subtracting $[n+t_jb_j+t_kb_k,m-t_ja_j-t_ka_k]$
in $\vA_i\vD_i$. This is because it can be easy verified that indeed
\begin{eqnarray}
&&\hspace{-.6in}(n+t_ib_i+t_jb_j)a_k+(m-t_ia_i-t_ja_j)b_k=\nonumber\\
&&\hspace{.4in}(n+t_ib_i+t_jb_j+t_kb_k)a_k
+(m-t_ia_i-t_ja_j-t_ka_k)b_k, \nonumber \\
&&\hspace{-.6in}(n+t_ib_i+t_kb_k)a_j+(m-t_ia_i-t_ka_k)b_j=\nonumber\\
&&\hspace{.4in}(n+t_ib_i+t_jb_j+t_kb_k)a_j
+(m-t_ia_i-t_ja_j-t_ka_k)b_j, \nonumber \\
&&\hspace{-.6in}(n+t_jb_j+t_kb_k)a_i+(m-t_ja_j-t_ka_k)b_i=\nonumber\\
&&\hspace{.4in}(n+t_ib_i+t_jb_j+t_kb_k)a_i
+(m-t_ia_i-t_ja_j-t_ka_k)b_i.
\end{eqnarray}

To eliminate all undesired elements we add to the vector in
(\ref{lc2}), $\binom {Q}{3}$  such columns, and the result is
\begin{eqnarray}
 \label{lc3}
&&\hspace{-.0in} \sum_{i=1}^{Q}[n+t_ib_i,m-t_ia_i]\exp( {j
{\omega_it_i}}) -\nonumber\\
&&\hspace{.0in}\sum_{i=1}^{Q-1}\sum_{j=i+1}^{Q}[n+t_ib_i+t_jb_j,m-t_ia_i-t_ja_j]\exp\left(
{j {[\omega_it_i+\omega_jt_j]}}\right)+
\nonumber \\
&&\hspace{-.0in}\sum_{i=1}^{Q-2}\sum_{j=i+1}^{Q-1}\sum_{k=j+1}^{Q}[n+t_ib_i+t_jb_j+t_kb_k,m-t_ia_i-t_ja_j-t_ka_k]\cdot
\nonumber\\
&&\hspace{1.1in}\cdot \exp\left( {j{[\omega_it_i+\omega_jt_j+\omega_kt_k]}}\right)
\end{eqnarray}

The last action canceled $\binom {Q}{2}(Q-2)=3\binom {Q}{3}$
undesired elements and created once again new $\binom
{Q}{3}(Q-3)=4\binom {Q}{4}$ undesired elements. We repeat this
procedure $Q$ times and in each step $k$, $1\leq k \leq Q$, we
subtract/add $\binom {Q}{k}$ columns for canceling $k\binom {Q}{k}$
undesired elements created in the previous step. Due to this
substraction/addition new $\binom {Q}{k}(Q-k)=(k+1)\binom {Q}{k+1}$
undesired elements are created. Clearly, when we subtract/add
$\binom {Q}{Q-1}$ columns exactly $Q=Q\binom {Q}{Q}$ undesired
elements are created. These may be canceled by substraction of a
single vector. By substraction/addition of the last vector,
$[n+t_1b_1+\ldots+t_Qb_Q,m-t_1a_1-\ldots-t_Qa_Q]$ the process
terminates, since we are canceling the last $Q$ undesired elements
and remain with $Q$ elements -- exactly those of the $[n,m]$ column,
\ie,
\begin{eqnarray}
 \label{lc4}
&& \hspace{-0.3in}[n,m]=\sum_{i=1}^{Q}[n+t_ib_i,m-t_ia_i]\exp( {j {\omega_it_i}})-\nonumber \\
&& \hspace{-0.3in}
\sum_{i=1}^{Q-1}\sum_{j=i+1}^{Q}[n+t_ib_i+t_jb_j,m-t_ia_i-t_ja_j]\exp\left(
{j  {[\omega_it_i+\omega_jt_j]}}\right)+\ldots
\nonumber \\
%&&\hspace{-0.3in}(-1)^{q-1}
%\mathop{\underbrace{\sum_{{i_1}=1}^{Q-q+1}\ldots\sum_{{i_q}={i_{q-1}}+1}^{Q}}}\limits_{\mbox{\emph{q}
%sums}}
%[n+t_{i_1}b_{i_1}+\ldots+t_{i_q}b_{i_q},m-t_{i_1}a_{i_1}-\ldots-t_{i_q}a_{i_q}]\cdot
%\nonumber \\
%&&\hspace{1.5in} \cdot \exp\left( {j
%{[\omega_{i_1}t_{i_1}+\ldots+\omega_{i_q}t_{i_q}]}}\right)+\ldots\nonumber
%\\
&&
\hspace{-0.3in}(-1)^{Q-1}[n+t_1b_1+\ldots+t_Qb_Q,m-t_1a_1-\ldots-t_Qa_Q]\exp\left(
{j {[\omega_1t_1+\ldots+\omega_Qt_Q]}}\right)=
\nonumber \\
&&\hspace{-0.3in} \sum_{q=1}^Q(-1)^{q-1}
\mathop{\underbrace{\sum_{{i_1}=1}^{Q-q+1}\ldots\sum_{{i_q}={i_{q-1}}+1}^{Q}}}\limits_{\mbox{\emph{q}
sums}}
[n+t_{i_1}b_{i_1}+\ldots+t_{i_q}b_{i_q},m-t_{i_1}a_{i_1}-\ldots-t_{i_q}a_{i_q}]\cdot
\nonumber \\
&&\hspace{1.5in} \cdot \exp\left( {j
{[\omega_{i_1}t_{i_1}+\ldots+\omega_{i_q}t_{i_q}]}}\right)
\end{eqnarray}

Clearly, this linear combination will be meaningful only if
$(t_1,\ldots,t_Q)\in T_Q^{(n,m)}$, \ie, $(n+t_{i_1}b_{i_1}+\ldots+
t_{i_q}b_{i_q},m-t_{i_1}a_{i_1}-\ldots-t_{i_q}a_{i_q})\in D,$ for
any $1\leq q \leq Q$, and where $(i_1,\ldots,i_q)$ is such that
$1\leq i_k \leq Q$ for all $1\leq k \leq q$.

%%%%%%%%%%%%%%%%%%%%%%%%%%%%%%%%%%%%%%%%%%%%%%%%%%%%%%
%
%       Bibliography
%
%%%%%%%%%%%%%%%%%%%%%%%%%%%%%%%%%%%%%%%%%%%%%%%%%%%%%%

%\*{\centerline{\bf Bibliography}}

\end{document}